\documentclass[%
 notitlepage,
 amsmath,amssymb,
 aps,
 reprint,
pra,
twocolumns,
nofootinbib
]{revtex4-1}

\usepackage[utf8]{inputenc}
\usepackage{graphicx}
\usepackage{hyperref}
\hypersetup{colorlinks = true, linkcolor = [rgb]{0.19411,0.51882,0.667058}, urlcolor = [rgb]{0.125490,0.29542,0.1647058}, citecolor = [rgb]{0.75882,0.37411,0.14117}}
\usepackage{amsmath}
\usepackage{blkarray}
\usepackage{amsfonts}
\usepackage{amssymb}
\usepackage{siunitx}
\usepackage{braket}
\usepackage{tikz}
\usepackage{import}
\usepackage{color,soul}
\newcommand*{\I}{\mathrm{i}\mkern1mu}
\usepackage[final]{changes}

\begin{document}

\title{Cumulative generation of maximal entanglement between spectrally distinct qubits using squeezed light}

\author{Elena Callus}\email{ecallus1@sheffield.ac.uk}
\author{Pieter Kok}\email{p.kok@sheffield.ac.uk}
\affiliation{Department of Physics and Astronomy, The University of Sheffield, Sheffield, S3 7RH, UK}

\begin{abstract}\noindent
We demonstrate how to create maximal entanglement between two qubits that are encoded in two spectrally distinct solid-state quantum emitters embedded in a waveguide interferometer. The optical probe is provided by readily accessible squeezed light, generated by parametric down-conversion. 
By continuously illuminating the emitters, the photon scattering and incremental path-erasure builds up entanglement. Our method does not require perfectly identical emitters, and accommodates spectral variations due to the fabrication process. \added{Furthermore, for some line-width and energy ratios, the entanglement build-up can be significantly faster than for more similar emitters.} It is also robust enough to create entanglement with a concurrence above 99\% in the event of scattering losses and detector inefficiencies, and can form the basis for practical entangled networks.
\end{abstract}

\date{\today}

\maketitle

\section{Introduction}
\noindent
A key resource for quantum computing and quantum information processing is entanglement 
\cite{Jozsa.2003}, \added{with the entanglement of bipartite systems a basic requirement for the implementation of a multi-partite quantum computer}. For photonic quantum technologies, entanglement can be generated between spatially separated solid-state emitters, or artificial atoms, by embedding them in wave-guides and allowing photons to interact with them. \added{The generation of entanglement between artificial atoms has been well-addressed theoretically, and various schemes to entangle such qubits already exist \cite{Cabrillo.1999,Plenio1999,Browne2003,Duan.2003,Simon2003,Feng2003,Barrett.2005,Metz2008,Yang2013,Li2016}. These processes require only single- or few-photon interactions and a relatively simple setup with few optical components, and utilise either controllable emitter interactions or ``which-path'' information erasure resulting from interference effects. In the case of the latter, some measurement is made that results in a state projection.}

\added{However, these schemes require the use of qubits that are spectrally identical for the coherent erasure of path information, and therefore any frequency variations between the emitters would result in degraded entanglement. Considering that current fabrication processes of solid-state emitters result in spectrally inhomogeneous samples \cite{Arakawa.2020}, the matching of sufficiently similar qubits adds a large overhead cost and the entanglement process cannot occur on a large scale. Although methods such as diameter tuning \cite{Heuser.2018}, strain tuning \cite{Zhai.2020} and utilising Raman transitions tuned into resonance \cite{Sipahigil2016}} can be used to tune the frequencies of the emitters, they require additional technical complexity in the experimental setup. Furthermore, these techniques are applicable only for sufficiently similar emitters and generally cannot be used for arbitrary pairings. 

In order to try and overcome this practical limitation, Hurst \textit{et al.} \cite{Hurst.2019} considered how spectral variation in emitters affects the entanglement outcome and demonstrated that this inhomogeneity is not as hindering as previously thought. They propose a simple setup involving linear optics and show that it is possible to attain entanglement deterministically for certain combinations of central energies and line-widths by adjusting the frequency of the probing photons. One drawback of the setup is that it requires the use of two-mode Fock states, $\ket{n,m}$, which are typically not easily accessible given current technologies. Also, for a given Fock state, deterministic near-perfect entanglement is attained only for certain ranges of central energy and line-width combinations. This necessitates the use of increasingly hard to source higher-order Fock states in order to entangle certain regimes of emitter-pairings.
\added{Levonian \textit{et al.} also consider the entanglement of emitters with well distinguishable central frequencies but reasonably closely matched line-widths \cite{Levonian2021}. This protocol makes use of interaction-free measurements for emitters in an interferometer, inspired by the Elitzur-Vaidman Gedanken experiment \cite{Aharonov2017}. 
}

In this paper, we show that two spectrally different emitters can be entangled with extremely high concurrence in a nearly deterministic manner by continuously illuminating the emitters with squeezed light, generated by means of spontaneous parametric down-conversion (SPDC) \cite{Mosley.2008}. This brings us one step closer to a physical implementation, given the accessibility of squeezed light \cite{Andersen.2016}, and further eases restrictions when it comes to the matching of inhomogeneous solid-state emitters. In addition, we take into consideration photon loss and show that very high concurrence is still possible in non-ideal situations. 

\added{This paper is organised as follows. In Sec. \ref{Protocol} we give a detailed presentation of the setup, as well as a theoretical framework for the scattering process and the generation of the squeezed state. In Sec. \ref{Entangling Proces} we study the state of the emitters as they evolve through the entanglement process, both in the lossless and the lossy case, and present our results using concurrence as the measure for entanglement. In Sec. \ref{Practical Implementation} we discuss physical implementation of the proposed scheme as well as practical challenges presently faced. In Sec. \ref{Conclusions} we review our findings and present our conclusions, and relevant derivations can be found in Appendixes A and B.}

\section{Protocol}\label{Protocol}

\begin{figure}[t!]
\includegraphics[width=\columnwidth]{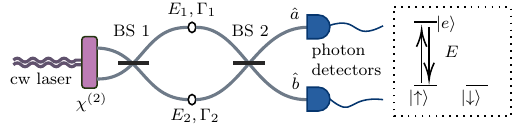}
\caption{Schematic of the setup: the squeezed light generated by a $\chi^{(2)}$ nonlinear crystal driven by a continuous-wave laser pump enters the Mach-Zehnder interferometer, where it scatters off the two solid-state emitters characterised by energies and line-widths $E_1$, $\Gamma_1$ and $E_2$, $\Gamma_2$. The photon measurement is made at the interferometer output, with the mode operators $\hat{a}$ and $\hat{b}$ representing the output arms. Inset shows the $L$-configuration of the emitters, with the $\ket{\uparrow}$ state coupled to the excited state, $\ket{e}$, and transition energy $E=E_1,\;E_2$.}
\label{fig:1}
\end{figure}

\added{In this section we introduce the setup of the proposed scheme, which consists} of a waveguide Mach-Zehnder interferometer (MZI) with a 50:50 beam splitter at either end, and solid-state emitters, acting as our logical qubits, embedded in each arm (see Fig.~\ref{fig:1}). The emitters have two long-lived low-lying spin states, $\ket{\uparrow}$ and $\ket{\downarrow}$, and an excited state $\ket{e}$, and are of the $L$-configuration, with the excited state coupled to only one of the spin states (say, the $\ket{\uparrow}$ state). The transition between the other spin state and the excited state is forbidden by polarisation selection rules. Each qubit is prepared in the superposition state $(\ket{\uparrow}+\ket{\downarrow})/\sqrt{2}$. To ensure that photons are scattered only in the forward direction, the two emitters are placed at so-called \emph{c-points} in the waveguide. These are the locations where emitters exhibit highly directional scattering of circularly polarized light due to the spatial confinement of the electromagnetic field \cite{Coles.2016}. High levels of directional scattering have been observed experimentally \cite{Coles.2016,Lodahl.2017,Lang.2016}. Finally, we have photon detectors placed at both output arms of the MZI, which herald the state of the emitters.

For emitters at c-points, the transmission coefficient for a photon of frequency $\omega$ scattering off a two-level emitter is obtained from the single-photon $S$-matrix, and is given by \cite{Shen.2007uyl}
\begin{equation}
t(\omega)=\dfrac{\hbar\omega-E-\I\hbar\left(\Gamma-\gamma\right)/2}{\hbar\omega-E+\I\hbar\left(\Gamma+\gamma\right)/2},
\end{equation}
where $E$ is the transition frequency of the emitter, and $\Gamma$ and $\gamma$ are the coupling rates of the emitter to the waveguide and the non-guided modes, respectively. The transmission coefficient of a photon that is lost to the environment is given by \cite{Rephaeli.2013}
\begin{align}
t_e(\omega)=\dfrac{-\I\hbar\sqrt{\Gamma\gamma}}{\hbar\omega-E+\I\hbar\left(\Gamma+\gamma\right)/2}.
\end{align}

In the case of zero photon loss, $\gamma=0$ and the scattered photon acquires a pure phase shift. In the ideal case, the two solid-state emitters are identical in their energies and line-widths and their spin states are entangled by passing a single resonant probe photon through the interferometer. After interfering with the first beam splitter, the photon scatters off the emitters and acquires a $\pi$ phase shift from its interaction with the $\ket{\uparrow}$ state \cite{Nysteen.2017,Fan.2010}, whilst the state $\ket{\downarrow}$ leaves the photon unchanged. The resulting state of the qubits and the probe photon after the second beam splitter is given by $(\ket{\Phi^-}\otimes\ket{1,0}-\ket{\Psi^-}\otimes\ket{0,1})/\sqrt{2}$, where $\ket{\Phi^-}=(\ket{\uparrow\uparrow}-\ket{\downarrow\downarrow})/\sqrt{2}$, $\ket{\Psi^-}=(\ket{\uparrow\downarrow}-\ket{\downarrow\uparrow})/\sqrt{2}$, and $\ket{1,0}$ and $\ket{0,1}$ represent the two possible photon-detection outcomes. Therefore, either photon measurement outcome would result in a maximally entangled bipartite state. However, if we allow for spectral variations between the two emitters, we do not obtain a maximally entangled state. Here, the amount of generated entanglement can be tuned by adjusting the frequency of the probe photon.

We consider the two-mode squeezed vacuum as our input state, routinely generated by spontaneous parametric down-conversion (SPDC) in a nonlinear $\chi^{(2)}$ crystal driven by a continuous-wave (CW) pump laser. During the SPDC process, a pump photon of frequency $\omega_p$ is annihilated and a signal and idler photon, with frequencies $\omega_s$ and $\omega_i$, respectively, are created. When acting on vacuum, the squeezing operator generates the state \cite{Kok.2010}
\begin{equation}
\ket{\psi}=\dfrac{1}{\cosh r}\sum^\infty_{n=0}(-\mathrm{e}^{\I\phi}\tanh r)^n\dfrac{(\hat{a}^\dagger\hat{b}^\dagger)^n}{n!}\ket{0},
\end{equation}
where $\hat{a}$, \smash{$\hat{b}$} are the mode operators for the two input arms of the MZI, and $\xi=r\mathrm{e}^{\I\phi}$ is determined by the material properties and the laser pump. We can ignore vacuum contributions since they do not affect the state of the multi-partite system in any way. We also neglect higher order photon pair production in the SPDC process as this occurs rarely; for typical experimental parameters and utilising a CW pump, the generation of multi-pair states as a fraction of single bi-photons is of the order of $10^{-8}$ per Watt of pump power \cite{Schneeloch.2019}. This places an upper limit to how strong the crystal can be pumped before multiple pair production changes the dynamics of the protocol. The low conversion efficiency in SPDC is due to the relative weakness of the signal and idler fields relative to the pump field \cite{Couteau.2018}.

\added{Two-photon scattering can be characterised by a linear term and a bound state term. For a given initial wave function $\beta(\omega_1,\omega_2)$, the post-scattering state $\tilde{\beta}(\omega_1,\omega_2)$ is expressed as \cite{Rephaeli.2013}}
\begin{widetext}
\begin{equation}\begin{split}
\tilde{\beta}_\alpha(\omega_1,\omega_2)=&\left[T(\omega_1,\omega_2)\beta(\omega_1,\omega_2)+T(\omega_2,\omega_1)\beta(\omega_2,\omega_1)\right]\\
&+\frac{\sqrt{\Gamma}}{\pi}S(\omega_1,\omega_2)\int\mathrm{d}k\left\lbrace\left[s(k)+s(\omega_1+\omega_2-k)\right]\beta(k,\omega_1+\omega_2-k)\right\rbrace,
\end{split}
\end{equation}
\end{widetext}
\added{where $\alpha$ indicates the number of photons lost to the environment. The first line of the expression relates to the (linear) single-photon scattering processes and depends solely on the initial wave function and the transmission coefficients of the individual photons. The second line is the bound state term and describes the frequency-mixing occurring during the multi-photon process. We define $T(\omega_1,\omega_2)$ and $S(\omega_1,\omega_2)$ as}
\begin{subequations}
\begin{equation}
T(\omega_1,\omega_2)=\begin{cases}
t(\omega_1)t(\omega_2)/2 \quad&\text{if }\alpha=0\\
t(\omega_1)t_e(\omega_2) \quad&\text{if }\alpha=1\\
t_e(\omega_1)t_e(\omega_2)/2 \quad&\text{if }\alpha=2
\end{cases}
\end{equation}
and
\begin{equation}
S(\omega_1,\omega_2)=\begin{cases}
\I s(\omega_1)s(\omega_2)/2 \quad&\text{if }\alpha=0\\
s(\omega_1)s_e(\omega_2) \quad&\text{if }\alpha=1\\
\I s_e(\omega_1)s_e(\omega_2)/2 \quad&\text{if }\alpha=2 ,
\end{cases}
\end{equation}
\end{subequations}
\added{where}
\begin{subequations}
\begin{equation}
s(\omega)=\frac{\hbar\sqrt{\Gamma}}{\hbar\omega-E+\I\hbar(\Gamma+\gamma)/2}
\end{equation}
and
\begin{equation}
s_e(\omega)=\frac{\hbar\sqrt{\gamma}}{\hbar\omega-E+\I\hbar(\Gamma+\gamma)/2}.
\end{equation}
\end{subequations}

We require the photons to be quasi-monochromatic, which can be achieved by either using a monochromatic pump beam, or by frequency filtering post-SPDC. For quasi-monochromatic photons, the scattering process can be described linearly, where the total phase shift accumulated during the interaction is the sum of the phase shifts imparted by the individual photons, \added{and the bound state term vanishes}. Additionally, photons with a broader bandwidth are more likely to excite the emitter due to their shorter temporal length, which may result in undesirable spontaneous emission \cite{Hurst.2019}.

In order to successfully reach maximal entanglement, the down-conversion process needs to be degenerate, i.e., producing signal and idler photons with the same frequency. A frequency difference between the signal and idler photons would impart which-path information during the scattering process, as the total acquired phase-shift is stronger for one emitter than the other (assuming non-identical emitters). The frequency of the generated photons needs to be optimised for the central energies and line-widths of the two quantum emitters. The phase shift imparted by the photons affects the interference at the second beam splitter and, consequently, the final state of our system. In order to successfully build up entanglement, we require $t^2_1(\omega)=t^2_2(\omega)$, where $t_i(\omega)$ is the transmission coefficient for the scattering in arm $i$. \added{We justify this condition in App. \ref{A}.}

\bigskip

\section{Entangling Process}\label{Entangling Proces}

\added{In this section, we present the state of our two-qubit system as it evolves through the entanglement generation process. We present both the zero photon-loss case as well as the lossy case, with losses occurring both during the scattering process and due to detector inefficiencies.}

\added{In both cases, we characterize the amount of entanglement for the two-qubit state post-photon detection, $\rho$, using the concurrence, $\mathcal{C}(\rho)$ \cite{Hill.1997}, which is given by}
\begin{equation}
\mathcal{C}(\rho)=\max(0,\lambda_1-\lambda_2-\lambda_3-\lambda_4),
\end{equation}
\added{where $\lambda_1,\ldots,\lambda_4$ are the eigenvalues, in decreasing order, of}
\begin{equation}
R=\sqrt{\sqrt{\rho}\left(\sigma_y\otimes\sigma_y\right)\rho^*\left(\sigma_y\otimes\sigma_y\right)\sqrt{\rho}}.
\end{equation}
\added{For a pure state, $\ket{\psi}=c_1\ket{\uparrow\uparrow}+c_2\ket{\uparrow\downarrow}+c_3\ket{\downarrow\uparrow}+c_4\ket{\downarrow\downarrow}$, the concurrence simplifies to}
\begin{equation}\begin{split}
\mathcal{C}\left(\ket{\psi}\right)&=|\braket{\psi|\sigma_y\otimes\sigma_y|\psi^*}|\\
&=2|c_1c_4-c_2c_3|,
\end{split}
\end{equation}
\added{and gives a measure of how much the state amplitudes deviate from those of a separable state. In the case of a mixed state, $\mathcal{C(\rho)}$ gives the average concurrence of a pure state decomposition of $\rho$, minimised over the different possible ensembles \cite{Wootters.2001}.}

\subsection{Lossless Case}

Consider the state of the system after $N$ detection events. Let $m$ be the number of events where both detectors register a photon and $n$ be the number of events where the two photons reach the same detector, with $N=m+n$. Then the state of the system after the $(N+1)^{\rm th}$ probe and right before photon-detection can be expressed as
\begin{widetext}
\begin{align}
\ket{\psi_{m,n}} =\; & \frac{1}{4c_{m,n}}\bigg[(1+t^2_1(\omega))^m(t^2_1(\omega)-1)^{n+1}\ket{\uparrow\downarrow}+(1+t^2_2(\omega))^m(1-t^2_2(\omega))^{n+1}\ket {\downarrow\uparrow}\bigg]\otimes\left[\left(\hat{a}^\dagger\right)^2+\left(\hat{b}^\dagger\right)^2\right]\ket{0} \cr
& +\dfrac{1}{2c_{m,n}}\bigg[ (t^2_1(\omega)+t^2_2(\omega))^{m+1}0^n\ket {\uparrow\uparrow}+(1+t^2_1(\omega))^{m+1}(t^2_1(\omega)-1)^{n}\ket {\uparrow\downarrow}\cr
& \qquad \qquad \quad +(1+t^2_2(\omega))^{m+1}(1-t^2_2(\omega))^{n}\ket{\downarrow\uparrow}+2^{m+1}0^n\ket{\downarrow\downarrow}\bigg]\otimes\hat{a}^\dagger\hat{b}^\dagger\ket{0}\, ,
\label{eq:statemain}
\end{align}
where $c_{m,n}$ is the respective normalization constant given by
\begin{equation}
c_{m,n}=\bigg[\left|(t^2_1(\omega)+t^2_2(\omega))^{m}(t^2_1(\omega)-t^2_2(\omega))^n\right|^2+\left|(1+t^2_1(\omega))^{m}(t^2_1(\omega)-1)^{n}\right|^2+\left|(1+t^2_2(\omega))^{m}(1-t^2_2(\omega))^{n}\right|^2+\left|2^{m}0^n\right|^2\bigg]^{1/2}.
\end{equation}
\end{widetext}
\added{We present a detailed derivation of the above in App. \ref{A}.}

Post-selecting on the photon detection, we obtain the heralded state of the emitters. Given that there is no information about which emitter has gained a phase shift, some of the which-path information of each photon is erased, resulting in a cumulative entanglement gain.

\begin{figure}[b]
\includegraphics[width=\columnwidth]{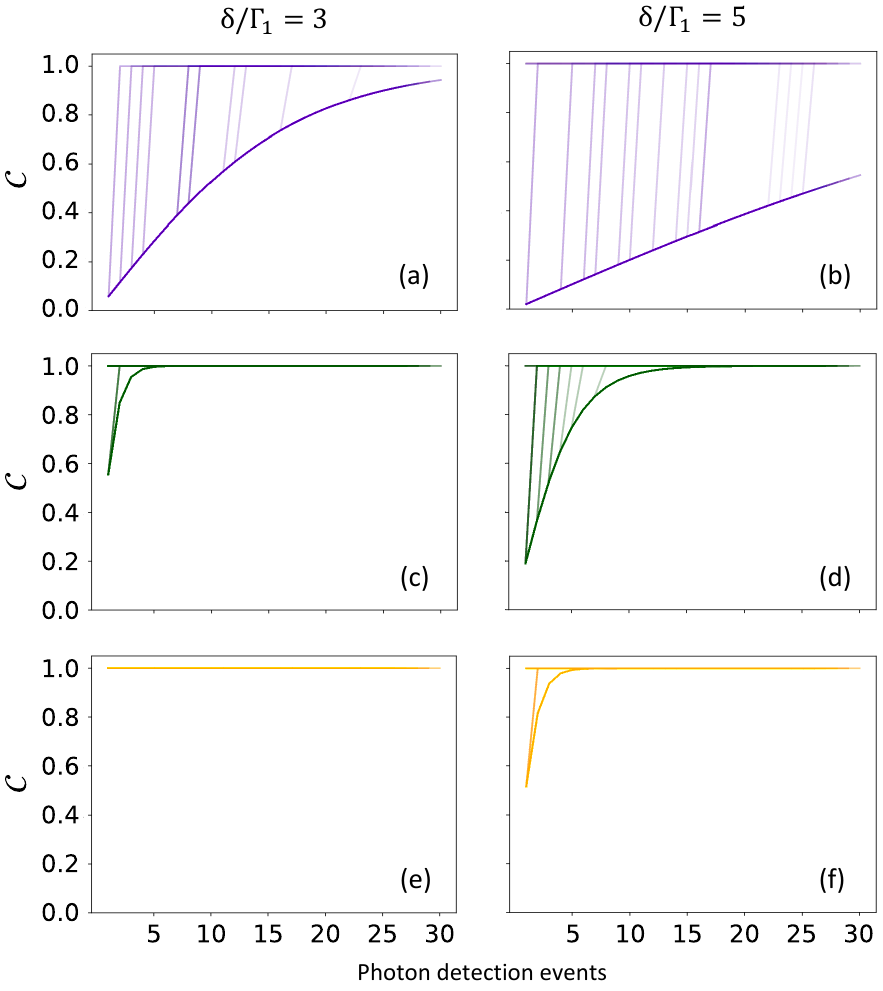}
\caption{Typical concurrence trajectories given a series of photon detection events (horizontal axis) for the lossless case ($\gamma=0$). $\Gamma_1$ and $\Gamma_2$ are the emitter line-widths, and the central energy detuning between the two emitters is $\delta=E_2-E_1$. Here, $\delta/\Gamma_1 =3$ (left column) and 5 (right column), and $\Gamma_2/\Gamma_1$ is set to 1 [(a) and (b)], 3 [(c) and (d)] and 5 [(e) and (f)]. We observe that within some spectral parameters, more dissimilar emitters can produce entanglement at a faster rate.}
\label{fig:2}
\end{figure}

\begin{figure*}[]
\includegraphics[width=\linewidth]{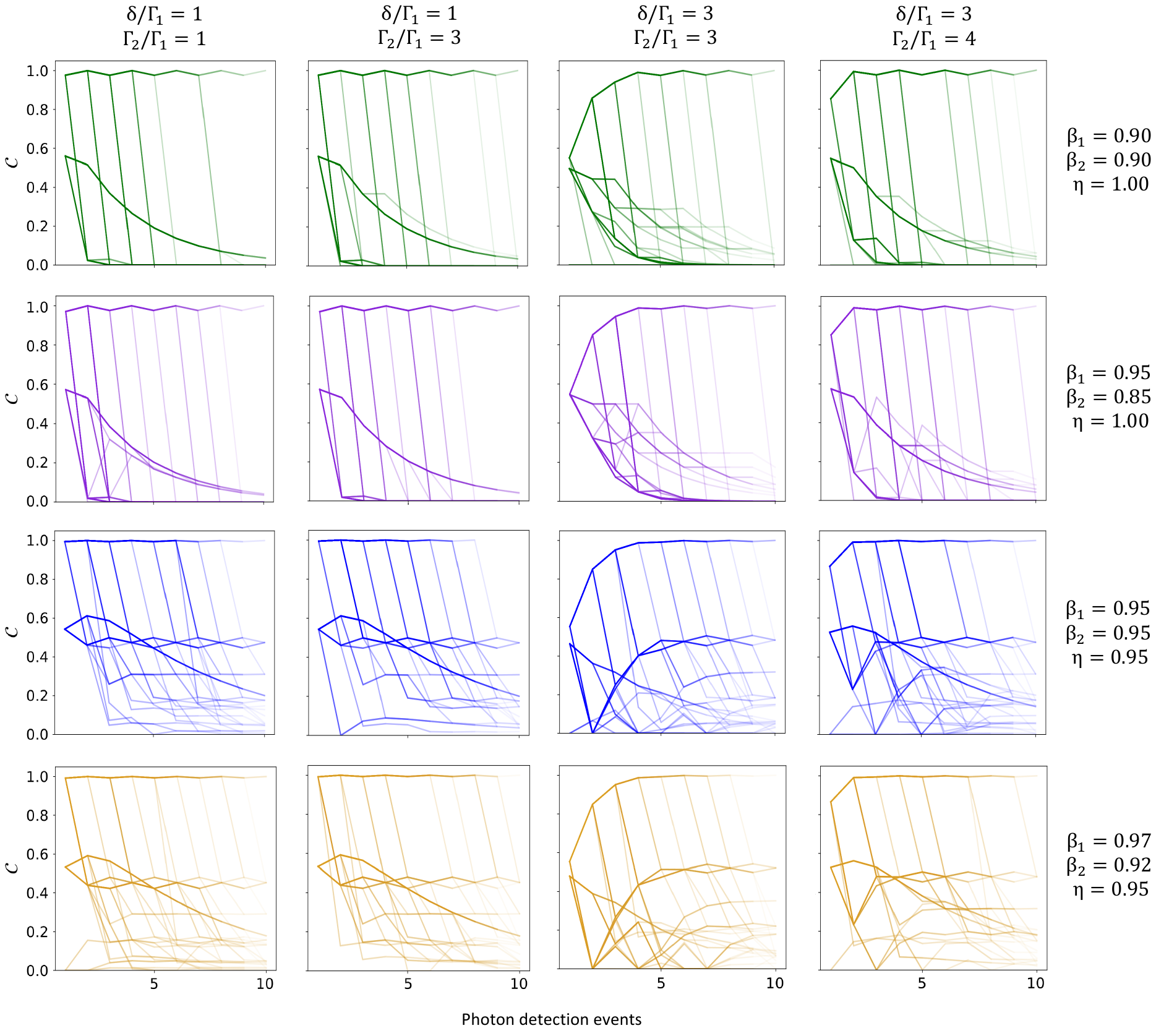}
\caption{Typical concurrence trajectories given a series of photon detection events for various $\beta$-factors, $\beta_1$ and $\beta_2$, and detector efficiencies, $\eta$. $\Gamma_1$ and $\Gamma_2$ are the emitter line-widths, and $\delta=E_2-E_1$ is the central energy detuning between the two emitters. The horizontal axis includes both successful photon-detection events and zero-photon detections due to scattering losses. The line transparency reflects the relative probability of a given path.}
\label{fig:3}
\end{figure*}

Fig.~\ref{fig:2} shows how the concurrence of two non-identical emitters can reach unity by repeatedly sending in photon pairs at the optimal frequency and keeping track of the photon detection signature.  Perfect entanglement is achieved regardless of the photon detection signature, and there is no need to reject samples on the basis of certain measurement outcomes. Furthermore, the process does not destroy any generated entanglement. Once both photons are detected by a single detector, the qubits are found to be in the maximally entangled Bell state $\ket{\Psi^\pm}=(\ket{\uparrow\downarrow}\pm\ket{\downarrow\uparrow})/\sqrt{2}$. Any further single-detector measurements simply toggle the state of the qubit between $\ket{\Psi^-}$ and $\ket{\Psi^+}$. Otherwise, if a coincidence measurement outcome, i.e. measurements where a photon reaches each detector, is registered at every iteration, the qubits approach the maximally entangled state $(\ket{\uparrow\uparrow}+\exp(\I\phi)\ket{\downarrow\downarrow})/\sqrt{2}$, where $\phi$ is a cumulative phase determined by $t_1(\omega)$ and $t_2(\omega)$.

The number of iterations needed to ensure that perfect entanglement has been reached is closely related to the expected value of concurrence after the first iteration, which in turn depends on the ratio of the line-widths of the two emitters as well as the detuning of their central energies. The trajectories in  Fig.~\ref{fig:2} show the random process of how the concurrence is updated after each photo-detection event. We see three possible paths a random trajectory can take. First, we have maximum entanglement from the first measurement, when both photons reach one detector. Second, we can have a slower build-up of entanglement, with the rate depending on the ratio of the energy detunings and line-widths. This occurs in the event of solely coincidence photon-measurements. Finally, we can have a single-detector click after a series of only coincident photon-measurement events, at which point the concurrence immediately goes to unity. The measurement process is characterised by a projection operator, assuming perfect photon counters. The scattering process is then repeated, where the state of the qubits is now the reduced density matrix after the previous measurement projection. Given a quasi-mono\-chro\-ma\-tic probe photon and zero photon loss, the (reduced) density matrix of the two emitters remains pure.
\newpage

In some situations, such as in Fig.~\ref{fig:2}, it may be desirable to pair emitters with larger linewidth variations in order to maximally entangle in fewer detection events without needing to consider the detector signatures. This is because the range of frequencies over which a scattering interaction results in a substantial phase shift can be improved by increasing the linewidth of the emitter. Therefore, it may be favourable to increase the linewidth of even just one emitter for a given energy detuning in order to improve the resulting phase shift. This in turn would result in smaller contributions to the $\ket{\uparrow\downarrow}$ and $\ket{\downarrow\uparrow}$ states in case of a coincidence measurement, heralding the entangled state $(\ket{\uparrow\uparrow}+\exp(\I\phi)\ket{\downarrow\downarrow})/\sqrt{2}$ at a faster rate.

In the case where the initial average concurrence is low, it is still possible to obtain maximal entanglement within fewer detection events, albeit with a lower probability, by considering the detector signatures; the concurrence reaches unity when both photons reach the same detector. This can be seen in the first line of Eq.~\eqref{eq:statemain}, where such an outcome would result in the Bell state $\ket{\Psi^\pm}=\smash{\frac{1}{\sqrt{2}}}(\ket{\uparrow\downarrow}\pm\ket{\downarrow\uparrow}$ (up to some overall phase). We find the remarkable property that more dissimilar emitters can produce entanglement at a faster rate. There are regimes of $\delta/\Gamma_1$ and $\Gamma_2/\Gamma_1$ where mere dissimilarity is not sufficient, but the range of emitters that can be entangled efficiently is vastly larger than when we require that all emitters are identical.

\subsection{Lossy Case}

We now consider photon loss during the scattering interaction at either emitter, which negatively impacts the entanglement process, as well as detector inefficiencies. The $\beta$-factor is defined as the coupling to the guided modes as a fraction of the total field coupling, $\beta={\Gamma}/{(\Gamma+\gamma)}$. We account for detector inefficiencies by considering a beam splitter with a transmission coefficient $\eta$ right before each photon detector.

Once we introduce scattering loss to our system, describing the scattering process becomes more involved as the photon in the guided mode will not only obtain a phase shift, but will also undergo a change in the probability amplitude (since $\gamma$ is no longer zero). \added{We include detailed description of the emitter density matrix in App. \ref{B}.}

One way to overcome the consequences of scattering losses is to consider photon number resolving detectors and discard samples where photon loss has occured. However, given that such detectors types are still in the experimental stage, we consider non-number resolving detectors in our calculations. The methodology for generating the trajectories is the same as for  $\beta=1$, but we will find that the qubits are in a mixed state post-photon detection. For emitter matches that reach perfect entanglement rapidly in the lossless case, it is possible to obtain over 99\% concurrence within a few iterations for $\beta\sim 0.9$.

In order to achieve the best concurrence, we include a bit flip operation on both emitters (in the computational basis) after every detection event. The scattering process results in an uneven accumulation of probability amplitudes on $\ket{\uparrow\uparrow}$ and $\ket{\downarrow\downarrow}$, and $\ket{\uparrow\downarrow}$ and $\ket{\downarrow\uparrow}$ due to $\gamma$ no longer being zero. Performing a bit flip balances out the probability amplitudes of these spin states and improves the amount of possible entanglement generation. Fig.~\ref{fig:3} shows random concurrence trajectories for the $\beta<1$ case. The different paths reflect the many different possible trajectories, with the probability weighting reflected by the degree of line transparency. Also, zero-photon detection outcomes due to scattering losses are accounted for. The plots demonstrate that the iterative process can generate near-perfect entanglement, where the probability of obtaining concurrence $C>0.99$ within the first 10 iterations is $10-50\%$ for the shown configurations. This occurs after several simultaneous photon measurements at both detectors. We find that there is no qualitative difference when allowing for non-identical $\beta$-factors at the two emitters.

\section{Practical Implementation}\label{Practical Implementation}

We now address challenges to the physical implementation of the proposed scheme. We have seen how photon loss degrades the entanglement generation process and decreases the rate at which samples can be successfully entangled. However, the results can be improved by implementing a bit-flip after every photon detection event. Alternatively, one can make use of photon-number-resolving detectors and discard samples where a photon is lost. 

\begin{figure}[h!]
\centering
\includegraphics[width=\columnwidth]{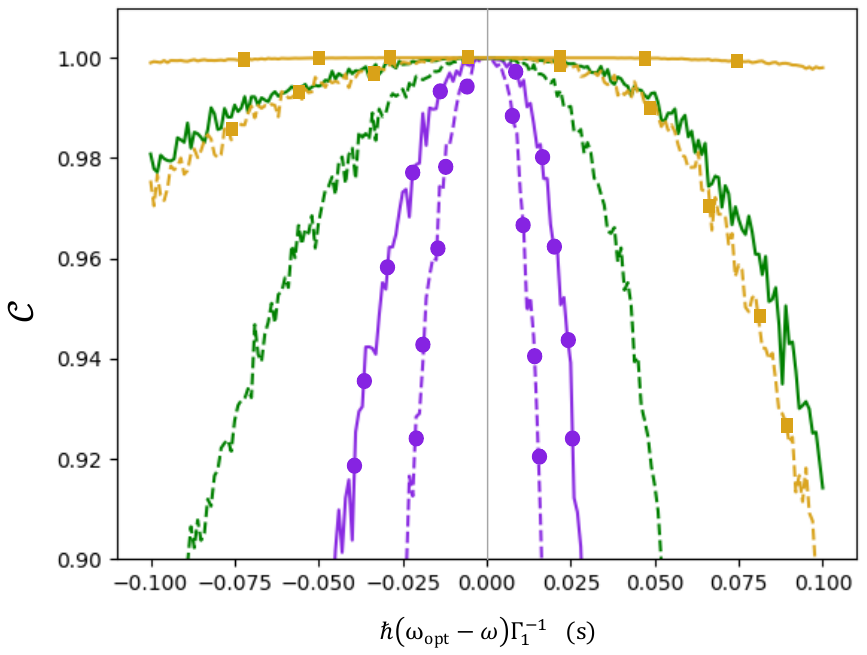}
\caption{The maximum possible concurrence, averaged over 300 trajectories, as a function of the difference between optimal frequency, $\omega_\text{opt}$, and actual input frequency, $\omega$. Here, $\Gamma_2/\Gamma_1$ is equal to 1 [purple, with circles], 3 [green, plain] and 5 [yellow, with squares], and $\delta/\Gamma_1$ is equal to 3 [solid] nd 5 [dotted].}
\label{fig:5}
\end{figure}

\added{One also needs to consider the impact of probe photons with frequencies that deviate from the optimum. In Fig.~\ref{fig:5} we plot the maximum possible concurrence, averaged over a large number of trajectories, against frequency deviation in the lossless case. If the input frequencies deviate slightly, this affects the amount of coherent interference at the second beam splitter stage and the extent of which-path information erasure. This in turn affects the build up of concurrence as the probing process is repeated, with the degree of impact depending on the energy detuning and linewidth ratio. For a quantum dot with linewidth of around \SI{1}{\micro e V}, the range of $(\omega_\text{opt}-\omega)$ plotted here corresponds to around $\pm 150$~MHz. The lack of symmetry in the plots arises from the fact that the rate of change of phase shift is not symmetric for frequencies slightly detuned from resonance.}

\begin{figure}[]
\includegraphics[width=\columnwidth]{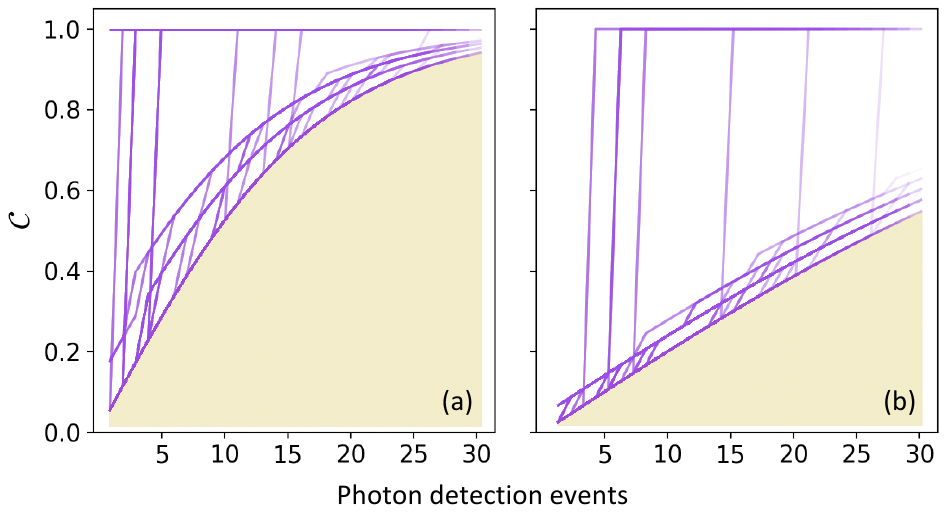}
\caption{Typical concurrence trajectories given a series of photon detection events (horizontal axis) for the lossless case ($\gamma=0$) when probing with both $\ket{1,1}$ and $\ket{2,2}$ states, where $\ket{2,2}$ states occur $\sim 15\%$ of the time. Here, $\delta/\Gamma_1 =3$ (a) and 5 (b), $\Gamma_2/\Gamma_1$ is set to 1, and the photon detectors are assumed to be number-resolving. The shaded region represents the area under the curve in Fig.~\ref{fig:2}. This shows that higher-order pair production in the down-conversion process is not detrimental to entanglement generation.}
\label{fig:4}
\end{figure}

Another physical issue that needs to be taken into account is the finite coherence time of the solid-state emitters, which is affected by mechanisms such as the spin-orbit and nuclear-spin interactions \cite{Fischer.2009,Coish.2009,Kloeffel.2013}. For semiconductor quantum dots, this coherence time is relatively short, ranging between $>100$ \si{\nano\second} and several microseconds \cite{Yoneda.2018,Prechtel.2016,Houel.2014,Petta.2005}. For nitrogen-vacancy centres in diamond, the coherence time can be in the millisecond range \cite{Herbschleb.2019,Zadrozny.2015,Jahnke.2012}, and exceeding half a second when enhanced by means of decoupling pulsing to suppress the spin decoherence \cite{Bar-Gill.2013}. The emitters must survive long enough for the entanglement generation to take place. This means that SPDC in the weak photon generation limit may be too slow. Strong pump amplitudes will create multiple pairs, however, and we must take into account the effect of four-photon scattering. Fig.~\ref{fig:4} compares the concurrence of the emitters when probing with just single biphoton states (i.e., $\ket{1,1}$) with possible higher-order pair production, using photon-number resolving detectors. The generation of multi-biphoton states does not disrupt the entanglement generation process, but rather may enhance it.

\section{Conclusions}\label{Conclusions}

In conclusion, we have presented a way to maximally entangle two solid-state quantum emitters via a cumulative entanglement generation protocol, taking into consideration the inhomogeneity arising from the fabrication process. Our protocol does not require that we discard entanglement due to undesired photon detection outcomes. We also have accounted for scattering photon losses and detector inefficiencies, and show that the results are still promising: for certain emitter pairings, it is still possible to create entanglement with a concurrence of $>0.99$ within a small number of photon-detection events. Additionally, the setup is relatively simple and can be implemented using current technology. We find that in trying to generate entanglement, we have more flexibility than previously thought; in fact, larger energy detuning or line-width variations might result in faster entanglement generation with higher concurrence, bringing us closer to solid state entanglement as a viable technology for quantum information processing.

\begin{acknowledgments}
The authors thank J.\ Iles-Smith and D.L.\ Hurst for valuable discussions. E.C.\ is supported by an EPSRC studentship. P.K.\ is supported by the EPSRC Quantum Communications Hub, Grant No. EP/M013472/1.
\end{acknowledgments}

\appendix
\section{Derivation of the Pure State in the Lossless Case}\label{A}

We derive here the post-measurement state of the qubit which, due to zero photon losses, remains pure regardless of the detection outcome. This is due to our choice of using a single frequency biphoton state and the absence of scattering losses -- each detection measurement corresponds to a single projection measurement. (Note that this no longer holds when using higher order biphoton states.) We therefore can make use of the usual wavefunction and ket notation to obtain a general expression for the state of the qubits after any number of photon-measurement events.

Following the same procedure as outlined in the supplemental information for \cite{Hurst.2019}, we start with both emitters in the state $\frac{1}{\sqrt{2}}(\ket{\uparrow}+\ket{\downarrow})$, giving an initial state
\begin{equation}
\ket{\psi_1}=\dfrac{1}{2}\left(\ket{\uparrow}+\ket{\downarrow}\right)\otimes\left(\ket{\uparrow}+\ket{\downarrow}\right)\otimes\hat{a}^\dagger\hat{b}^\dagger\ket{0},
\end{equation}
where $\hat{a}^\dagger$ and $\hat{b}^\dagger$ are the creation operators for the upper and lower input arms of the interferometer, both with frequency $\omega$. 

The state after the first beam-splitter ($\hat{a}\rightarrow\frac{1}{\sqrt{2}}\left(\hat{a}+\hat{b}\right)$ and $\hat{b}\rightarrow\frac{1}{\sqrt{2}}\left(\hat{a}-\hat{b}\right)$) is
\begin{equation}\begin{split}
\ket{\psi_2}=\;&\dfrac{1}{4}\left(\ket{\uparrow}+\ket{\downarrow}\right)\otimes\left(\ket{\uparrow}+\ket{\downarrow}\right)\otimes\left(\hat{a}^\dagger+\hat{b}^\dagger\right)\left(\hat{a}^\dagger-\hat{b}^\dagger\right)\ket{0}\\
=\;&\dfrac{1}{4}\left(\ket{\uparrow}+\ket{\downarrow}\right)\otimes\left(\ket{\uparrow}+\ket{\downarrow}\right)\otimes\left[\left(\hat{a}^\dagger\right)^2-\left(\hat{b}^\dagger\right)^2\right]\ket{0}.\\
\end{split}
\end{equation}

The state post-scattering is then
\begin{equation}\begin{split}
\ket{\psi_3}=\; & \frac{1}{4}\left(t_1^2(\omega)\ket{\uparrow}+\ket{\downarrow}\right)\otimes\left(\ket{\uparrow}+\ket{\downarrow}\right)\otimes\left(\hat{a}^\dagger\right)^2\ket{0}\\
&-\frac{1}{4}\left(\ket{\uparrow}+\ket{\downarrow}\right)\otimes\left(t^2_2(\omega)\ket{\uparrow}+\ket{\downarrow}\right)\otimes\left(\hat{b}^\dagger\right)^2\ket{0}.
\end{split}
\end{equation}

The state after the second beam-splitter is given by
\begin{widetext}
\begin{equation}\begin{split}
\ket{\psi}=\;&\frac{1}{8}\bigg[(t^2_1(\omega)-t^2_2(\omega))\ket{\uparrow\uparrow}+(t^2_1(\omega)-1)\ket{\uparrow\downarrow}+(1-t^2_2(\omega))\ket{\downarrow\uparrow}+0\ket{\downarrow\downarrow}\bigg]\otimes\bigg[\left(\hat{a}^\dagger\right)^2+\left(\hat{b}^\dagger\right)^2\bigg]\ket{0}\\
&+\frac{1}{4}\bigg[(t^2_1(\omega)+t^2_2(\omega))\ket{\uparrow\uparrow}+(t^2_1(\omega)+1)\ket{\uparrow\downarrow}+(1+t^2_2(\omega))\ket{\downarrow\uparrow}+2\ket{\downarrow\downarrow}\bigg]\otimes\hat{a}^\dagger\hat{b}^\dagger\ket{0}.
\end{split}
\label{eq:state}
\end{equation}
\end{widetext}

Making use of projection measurements once again, a measurement outcome heralds one of the following states:
\begin{widetext}
\begin{subequations}
\begin{equation}\begin{split}
\ket{\psi_{(1,0)}}=\ket{\psi_{(0,1)}}=\;&\frac{(t^2_1(\omega)-t^2_2(\omega))\ket{\uparrow\uparrow}+(t^2_1(\omega)-1)\ket{\uparrow\downarrow}+(1-t^2_2(\omega))\ket{\downarrow\uparrow}+0\ket{\downarrow\downarrow}}{\sqrt{\left[|t^2_1(\omega)-t^2_2(\omega)|^2+|t^2_1(\omega)-1|^2+|1-t^2_2(\omega)|^2\right]}}
\end{split}
\end{equation}
\begin{equation}\begin{split}
\text{or} \qquad \ket{\psi_{(1,1)}}=\;&\frac{(t^2_1(\omega)+t^2_2(\omega))\ket{\uparrow\uparrow}+(t^2_1(\omega)+1)\ket{\uparrow\downarrow}+(1+t^2_2(\omega))\ket{\downarrow\uparrow}+2\ket{\downarrow\downarrow}}{\sqrt{\left[|t^2_1(\omega)+t^2_2(\omega)|^2+|t^2_1(\omega)+1|^2+|1+t^2_2(\omega)|^2+4\right]}}.
\end{split}
\end{equation}
\end{subequations}
\end{widetext}
We then repeat the probing process and replace the initial state of the emitters with either of the heralded states. 

Let us consider the system after $N=m+n$ detection events, where $m$ is the number of events where both detectors register a photon and $n$ be the number of events where the two photons reach the same detector. Then we can express the state after the $(N+1)^{\rm th}$ probe and right before photon-detection as
\begin{widetext}
\begin{equation}\begin{split}
\ket{\psi_{m,n}} =\; & \frac{1}{4c_{m,n}}\bigg[(t^2_1(\omega)+t^2_2(\omega))^m(t^2_1(\omega)-t^2_2(\omega))^{n+1}\ket{\uparrow\uparrow}+(1+t^2_1(\omega))^m(t^2_1(\omega)-1)^{n+1}\ket{\uparrow\downarrow}\\
&\qquad \qquad \quad +(1+t^2_2(\omega))^m(1-t^2_2(\omega))^{n+1}\ket {\downarrow\uparrow}\bigg]\otimes\left[\left(\hat{a}^\dagger\right)^2+\left(\hat{b}^\dagger\right)^2\right]\ket{0} \\
& +\dfrac{1}{2c_{m,n}}\bigg[ (t^2_1(\omega)+t^2_2(\omega))^{m+1}(t_1^2(\omega)-t^2_2(\omega))^n\ket {\uparrow\uparrow}+(1+t^2_1(\omega))^{m+1}(t^2_1(\omega)-1)^{n}\ket {\uparrow\downarrow}\\
& \qquad \qquad \quad +(1+t^2_2(\omega))^{m+1}(1-t^2_2(\omega))^{n}\ket{\downarrow\uparrow}+2^{m+1}0^n\ket{\downarrow\downarrow}\bigg]\otimes\hat{a}^\dagger\hat{b}^\dagger\ket{0}\, 
\end{split}
\label{eq:rec}
\end{equation}
where $c_{m,n}$ is the normalization constant given by
\begin{equation}
c_{m,n}=\bigg[\left|(t^2_1(\omega)+t^2_2(\omega))^{m}(t^2_1(\omega)-t^2_2(\omega))^n\right|^2+\left|(1+t^2_1(\omega))^{m}(t^2_1(\omega)-1)^{n}\right|^2+\left|(1+t^2_2(\omega))^{m}(1-t^2_2(\omega))^{n}\right|^2+\left|2^{m}0^n\right|^2\bigg]^{1/2}.
\end{equation}
\end{widetext}

Next we justify the limitation placed on the choice of frequency: by selecting a frequency that satisfies $t^2_1(\omega)=t^2_2(\omega)\equiv t^2(\omega)$, \eqref{eq:rec} simplifies to
\begin{widetext}
\begin{equation}\begin{split}
\ket{\psi_{m,n}} =\; & \frac{1}{4c_{m,n}}\bigg[(1+t^2(\omega))^m(t^2(\omega)-1)^{n+1}\ket{\uparrow\downarrow}+(1+t^2(\omega))^m(1-t^2(\omega))^{n+1}\ket{\downarrow\uparrow}\bigg]\otimes\left[\left(\hat{a}^\dagger\right)^2+\left(\hat{b}^\dagger\right)^2\right]\ket{0} \\
& +\dfrac{1}{2c_{m,n}}\bigg[ (t^2(\omega)+t^2(\omega))^{m+1}0^n\ket {\uparrow\uparrow}+(1+t^2(\omega))^{m+1}(t^2(\omega)-1)^{n}\ket {\uparrow\downarrow}\\
& \qquad \qquad \quad +(1+t^2(\omega))^{m+1}(1-t^2(\omega))^{n}\ket{\downarrow\uparrow}+2^{m+1}0^n\ket{\downarrow\downarrow}\bigg]\otimes\hat{a}^\dagger\hat{b}^\dagger\ket{0}.
\end{split}
\end{equation}
\end{widetext}
Therefore, the choice of frequency allows us to obtain a maximally entangled Bell state $\ket{\Psi^\pm}=(\ket{\uparrow\downarrow}+(-1)^{n+1}\ket{\downarrow\uparrow})/\sqrt{2}$ (up to some global phase) when both photons reach just one detector, or to approach the maximally entangled state $\exp(\I\phi)\ket{\uparrow\uparrow}+\ket{\downarrow\downarrow}$. An expression for the relative phase can be obtained by considering how long it takes for the emitters to reach this state in the event of only coincident detections: if this state is reached after $M$ consecutive coincident photon-detections, then $\exp(\I\phi)=t^{2M}_1(\omega)=t^{2M}_2(\omega).$

The frequency $\omega$ that satisfies the condition $t^2_1(\omega)=t^2_2(\omega)$ is given by either of the following:
\begin{subequations}
\begin{equation}
\hbar \omega=\dfrac{1}{2}\left[E_1+E_2\pm\sqrt{\left(E_1-E_2\right)^2-\hbar^2\Gamma_1 \Gamma_2}\right] ,
\end{equation}
\begin{equation}
\hbar \omega=\frac{E_2\Gamma_1-E_1\Gamma_2}{\Gamma_1-\Gamma_2} .
\end{equation}
\label{eq:optimal}
\end{subequations}

\section{Derivation of the Density Matrix in the Lossy Case}\label{B}

We now consider possible losses during our entanglement process and express the state of the two emitters using the density matrix formalism (this is due to the possibility of obtaining a mixed state post-photon detection). We choose a frequency that satisfies Eq.~\eqref{eq:optimal}, where $t_1(\omega)=\pm t_2(\omega)$ no longer holds since $\gamma$ is no longer zero. Furthermore, it may be the case that the three frequency choices result in different scattering amplitudes for a given emitter pairing, and therefore, in different trajectories. We suppress the scattering amplitude notation so that $t_i(\omega)\rightarrow t_i$ and $t_{e,i}(\omega)\rightarrow t_{e,i}$.

We start off with an arbitrary density matrix for the qubits, $\rho_\text{emitters}$, and a photon in each input arm of the interferometer, $\hat{a}$ and $\hat{b}$:
\begin{equation}
\rho_1=\rho_\text{emitters}\otimes\big[\hat{a}^\dagger\hat{b}^\dagger\ket{0}\bra{0}\hat{a}\,\hat{b}\big].
\end{equation}
For the first iteration,
\begin{equation}
\rho_\text{emitters}=\frac{1}{4}\left[\left(\ket{\uparrow}+\ket{\downarrow}\right)\otimes\left(\ket{\uparrow}+\ket{\downarrow}\right)\right]\left[\text{H.c.}\right],
\label{eq:rhoemitters}
 \end{equation}
where H.c. is the Hermitian conjugate.

Interacting with the first beamsplitter, the state evolves to
\begin{equation}
\rho_2=\rho_\text{emitters}\otimes\frac{1}{4}\bigg[\left(\hat{a}^\dagger\right)^2-\left(\hat{b}^\dagger\right)^2\bigg]\ket{0}\bra{0}\big[\hat{a}^2-\hat{b}^2\big].
\end{equation}

The state after the scattering interactions at the emitters is
\begin{widetext}
\begin{equation}
\begin{split}
\rho_3=&\;\frac{1}{4}\sum_{i,j= 1,2,3} \rho(M_{i},M_{j})\otimes\big[M_{i}\ket{0}\bra{0}M_{j}\big]+\frac{1}{4}\sum_{i,j= 1,2,3} \rho(N_{i},N_{j})\otimes\big[N_{i}\ket{0}\bra{0}N_{j}\big]\\
&-\frac{1}{4}\sum_{i,j=1,2,3} \left\lbrace\rho(M_{i},N_{j})\otimes\big[M_{i}\ket{0}\bra{0}N_{j}\big]+\rho(N_{i},M_{j})\otimes\big[N_{i}\ket{0}\bra{0}M_{j}\big]\right\rbrace,
\end{split}
\end{equation}
\end{widetext}
where
\begin{subequations}
\begin{equation}
M=\;\begin{blockarray}{c}
\begin{block}{[c]}
   \hat{a}^2 \\
   \hat{a}\hat{r}_1  \\
   \hat{r}_1^2 \\
\end{block}
\end{blockarray}~,
 \end{equation}
 \begin{equation}
N=\;\begin{blockarray}{cc}
\begin{block}{[cc]}
   \hat{b}^2 \\
   \hat{b}\hat{r}_2 \\
   \hat{r}_2^2 \\
\end{block}
\end{blockarray}~,
 \end{equation}
 \end{subequations}
and where $\hat{r}_i$ is the photon annihilation operator to a reservoir around the emitter in arm $i$ which represents scattering losses, and $\rho(\hat{m},\hat{n})$ is the respective density matrix of the qubits associated with the scattered optical state $\hat{m}^\dagger\ket{0}\bra{0}\hat{n}$. The scattering amplitudes are given by the following transformations:
\begin{widetext}
\begin{subequations}
\begin{equation}
\begin{pmatrix}
   \ket{\uparrow\uparrow} \\
   \ket{\uparrow\downarrow} \\
   \ket{\downarrow\uparrow} \\
   \ket{\downarrow\downarrow} \\
\end{pmatrix}\otimes\left(\hat{a}^\dagger\right)^2\ket{0}\longrightarrow
\begin{pmatrix}
   t_1^2\ket{\uparrow\uparrow} \\
   t_1^2\ket{\uparrow\downarrow} \\
   \ket{\downarrow\uparrow} \\
   \ket{\downarrow\downarrow} \\
\end{pmatrix}\otimes\left(\hat{a}^\dagger\right)^2\ket{0}+
\begin{pmatrix}
   2t_1 t_{e,1}\ket{\uparrow\uparrow} \\
   2t_1 t_{e,1}\ket{\uparrow\downarrow} \\
   0 \\
   0 \\
\end{pmatrix}\otimes\hat{a}^\dagger\hat{r}^\dagger_1\ket{0}+
\begin{pmatrix}
   t_{e,1}^2\ket{\uparrow\uparrow} \\
   t_{e,1}^2\ket{\uparrow\downarrow} \\
   0 \\
   0 \\
\end{pmatrix}\otimes\left(\hat{r}_1^\dagger\right)^2\ket{0},
 \end{equation}
\begin{equation}
\begin{pmatrix}
   \ket{\uparrow\uparrow} \\
   \ket{\uparrow\downarrow} \\
   \ket{\downarrow\uparrow} \\
   \ket{\downarrow\downarrow} \\
\end{pmatrix}\otimes\left(\hat{b}^\dagger\right)^2\ket{0}\longrightarrow
\begin{pmatrix}
   t_2^2\ket{\uparrow\uparrow} \\
   \ket{\uparrow\downarrow} \\
   t_2^2\ket{\downarrow\uparrow} \\
   \ket{\downarrow\downarrow} \\
\end{pmatrix}\otimes\left(\hat{b}^\dagger\right)^2\ket{0}+
\begin{pmatrix}
   2t_2 t_{e,2}\ket{\uparrow\uparrow} \\
   0 \\
   2t_2 t_{e,2}\ket{\downarrow\uparrow} \\
   0 \\
\end{pmatrix}\otimes\hat{b}^\dagger\hat{r}^\dagger_2\ket{0}+
\begin{pmatrix}
   t_{e,1}^2\ket{\uparrow\uparrow} \\
   0 \\
   t_{e,2}^2\ket{\downarrow\uparrow} \\
   0 \\
\end{pmatrix}\otimes\left(\hat{r}_2^\dagger\right)^2\ket{0}.
 \end{equation}
\end{subequations}

The state then interacts with the second beamsplitter, leaving the state of the qubits unaltered and only changing the optical state:
\begin{equation}
\begin{split}
\rho_4=&\;\frac{1}{4}\sum_{i,j= 1,2,3} \rho(M_{i1},M_{j1})\otimes\big[M_{i,2}\ket{0}\bra{0}M_{j2}\big]+\frac{1}{4}\sum_{i,j= 1,2,3} \rho(N_{i1},N_{j1})\otimes\big[N_{i,2}\ket{0}\bra{0}N_{j,2}\big]\\
&-\frac{1}{4}\sum_{i,j=1,2,3} \left\lbrace\rho(M_{i1},N_{j1})\otimes\big[M_{i2}\ket{0}\bra{0}N_{j2}\big]+\rho(N_{i1},M_{j1})\otimes\big[N_{i2}\ket{0}\bra{0}M_{j2}\big]\right\rbrace,
\end{split}
\label{eq:rho}
\end{equation}
\end{widetext}
where $M$ and $N$ now change to
\begin{subequations}
\begin{equation}
M=\;\begin{blockarray}{cc}
\begin{block}{[cc]}
   \hat{a}^2 & \frac{1}{2}(\hat{a}^2+2\hat{a}\hat{b}+\hat{b}^2) \\
   \hat{a}\hat{r}_1 & \frac{1}{\sqrt{2}}(\hat{a}+\hat{b})\hat{r}_1  \\
   \hat{r}_1^2 & \hat{r}_1^2  \\
\end{block}
\end{blockarray}~,
 \end{equation}
 \begin{equation}
N=\;\begin{blockarray}{cc}
\begin{block}{[cc]}
   \hat{b}^2 & \frac{1}{2}(\hat{a}^2-2\hat{a}\hat{b}+\hat{b}^2) \\
   \hat{b}\hat{r}_2 & \frac{1}{\sqrt{2}}(\hat{a}-\hat{b})\hat{r}_2  \\
   \hat{r}_2^2 & \hat{r}_2^2  \\
\end{block}
\end{blockarray}~.
 \end{equation}
 \end{subequations}
 
Finally, we consider photon detector inefficiencies by considering a beam splitter with transmission coefficient $\eta$ before each detector, such that $\hat{a}\rightarrow\left(\sqrt{\eta}\,\hat{a}+\sqrt{1-\eta}\,\hat{r}_3\right)$ and $\hat{b}\rightarrow\left(\sqrt{\eta}\,\hat{b}+\sqrt{1-\eta}\,\hat{r}_4\right)$. This step leaves the state of the qubits unchanged, with the optical state changing as follows:
\begin{widetext}
\begin{equation}
\begin{split}
\rho_5=&\;\frac{1}{4}\sum_{i,j= 1,2,3} \rho(M_{i1},M_{j1})\otimes\big[M_{i,2}\ket{0}\bra{0}M_{j2}\big]+\frac{1}{4}\sum_{i,j= 1,2,3} \rho(N_{i1},N_{j1})\otimes\big[N_{i,2}\ket{0}\bra{0}N_{j,2}\big]\\
&-\frac{1}{4}\sum_{i,j=1,2,3} \left\lbrace\rho(M_{i1},N_{j1})\otimes\big[M_{i2}\ket{0}\bra{0}N_{j2}\big]+\rho(N_{i1},M_{j1})\otimes\big[N_{i2}\ket{0}\bra{0}M_{j2}\big]\right\rbrace,
\end{split}
\label{eq:rho}
\end{equation}
where $M$ and $N$ now change to
\begin{subequations}
\begin{equation}
M=\;\begin{blockarray}{cc}
\begin{block}{[cc]}
   \hat{a}^2 & \frac{1}{2}\left[(\sqrt{\eta}\,\hat{a}+\sqrt{1-\eta}\,\hat{r}_3)^2+(\sqrt{\eta}\,\hat{b}+\sqrt{1-\eta}\,\hat{r}_4)^2+2(\sqrt{\eta}\,\hat{a}+\sqrt{1-\eta}\,\hat{r}_3)(\sqrt{\eta}\,\hat{b}+\sqrt{1-\eta}\,\hat{r}_4)\right] \\
   \hat{a}\hat{r}_1 & \frac{1}{\sqrt{2}}(\sqrt{\eta}\,\hat{a}+\sqrt{\eta}\,\hat{b}+\sqrt{1-\eta}\,\hat{r}_3+\sqrt{1-\eta}\,\hat{r}_4)\hat{r}_1  \\
   \hat{r}_1^2 & \hat{r}_1^2  \\
\end{block}
\end{blockarray}~,
 \end{equation}
 \begin{equation}
N=\;\begin{blockarray}{cc}
\begin{block}{[cc]}
   \hat{b}^2 & \frac{1}{2}\left[(\sqrt{\eta}\,\hat{a}+\sqrt{1-\eta}\,\hat{r}_3)^2+(\sqrt{\eta}\,\hat{b}+\sqrt{1-\eta}\,\hat{r}_4)^2-2(\sqrt{\eta}\,\hat{a}+\sqrt{1-\eta}\,\hat{r}_3)(\sqrt{\eta}\,\hat{b}+\sqrt{1-\eta}\,\hat{r}_4)\right] \\
   \hat{b}\hat{r}_2 & \frac{1}{\sqrt{2}}(\sqrt{\eta}\,\hat{a}-\sqrt{\eta}\,\hat{b}+\sqrt{1-\eta}\,\hat{r}_3-\sqrt{1-\eta}\,\hat{r}_4)\hat{r}_2  \\
   \hat{r}_2^2 & \hat{r}_2^2  \\
\end{block}
\end{blockarray}~.
 \end{equation}
 \end{subequations}
 \end{widetext}
 
We make use of positive operator-valued measures made up of orthogonal projectors, where $\Pi(\hat{x})=\hat{x}^\dagger\ket{0}\bra{0}\hat{x}$, and assume non-photon-number resolving detectors. 

A click by the detector at output arm $\hat{a}$, with no detection in the other detector, is only possible for the projection measurements $\Pi(\hat{a}^2)$, $\Pi(\hat{a}\hat{r}_1)$ and $\Pi(\hat{a}\hat{r}_2)$. This yields the following reduced density matrix for the qubits:
\begin{widetext}
\begin{equation}\begin{split}
\rho_{(1,0)}&=\frac{1}{P(1,0)}\text{Tr}_\text{field}\left[\left(\Pi(\hat{a}^2)+\Pi(\hat{a}\hat{r}_1)+\Pi(\hat{a}\hat{r}_2)+\Pi(\hat{a}\hat{r}_3)+\Pi(\hat{a}\hat{r}_4)\right)\rho_5\left(\Pi(\hat{a}^2)+\Pi(\hat{a}\hat{r}_1)+\Pi(\hat{a}\hat{r}_2)+\Pi(\hat{a}\hat{r}_3)+\Pi(\hat{a}\hat{r}_4)\right)\right]\\
&=\frac{1}{P(1,0)}\bigg\lbrace\frac{\eta^2}{16}\left[\rho\left(\hat{a}^2,\hat{a}^2\right)+\rho\left(\hat{b}^2,\hat{b}^2\right)-\rho\left(\hat{a}^2,\hat{b}^2\right)-\rho\left(\hat{b}^2,\hat{a}^2\right)\right]\\
&\qquad \qquad \qquad \qquad+\frac{\eta(1-\eta)}{2}\left[\rho\left(\hat{a}^2,\hat{a}^2\right)+\rho\left(\hat{b}^2,\hat{b}^2\right)\right]+\frac{\eta}{8}\left[\rho(\hat{a}\hat{r}_1,\hat{a}\hat{r}_1)+\rho(\hat{b}\hat{r}_2,\hat{b}\hat{r}_2)\right] \bigg\rbrace.\\
\end{split}
\label{eq:rho10}
\end{equation} 

The probability of obtaining this measurement outcome, $P(1,0)$, is
\begin{equation}\begin{split}
P(1,0)&=\text{Tr}\left[\left(\Pi(\hat{a}^2)+\Pi(\hat{a}\hat{r}_1)+\Pi(\hat{a}\hat{r}_2)+\Pi(\hat{a}\hat{r}_3)+\Pi(\hat{a}\hat{r}_4)\right)\rho_5\left(\Pi(\hat{a}^2)+\Pi(\hat{a}\hat{r}_1)+\Pi(\hat{a}\hat{r}_2)+\Pi(\hat{a}\hat{r}_3)+\Pi(\hat{a}\hat{r}_4)\right)\right].\\
\end{split}
\label{eq:P10}
\end{equation}
\end{widetext}

For the measurement outcome $(0,1)$, we use the projectors $\Pi(\hat{b}^2)$, $\Pi(\hat{b}\hat{r}_1)$, $\Pi(\hat{b}\hat{r}_2)$, $\Pi(\hat{b}\hat{r}_3)$ and $\Pi(\hat{b}\hat{r}_4)$, which gives the same result as for the previous case, i.e., $\rho_{(0,1)}$ is given by Eq.~\eqref{eq:rho10} and $P(0,1)$ is given by Eq.~\eqref{eq:P10}.

In the case of a coincidence detection, the post-measurement density matrix is given by
\begin{widetext}
\begin{equation}\begin{split}
\rho_{(1,1)}&=\frac{1}{P(1,1)}\text{Tr}_\text{field}\left[\Pi(\hat{a}\hat{b})\rho_4\Pi(\hat{a}\hat{b})\right]=\frac{\eta^2}{4P(1,1)}\left[\rho(\hat{a}^2,\hat{a}^2)+\rho(\hat{b}^2,\hat{b}^2)+\rho(\hat{a}^2,\hat{b}^2)+\rho(\hat{b}^2,\hat{a}^2)\right],\\
\end{split}
\end{equation}
\end{widetext}
where 
\begin{equation}
\begin{split}
P(1,1)&=\text{Tr}\left[\Pi(\hat{a}\hat{b})\rho_4\Pi(\hat{a}\hat{b})\right].\\
\end{split}
\end{equation}
Note that this reduced matrix represents a pure state.

Finally, we have the case of no photon detection, represented by the following state:
\begin{widetext}
\begin{equation}
\begin{split}
\rho_{(0,0)}&=\frac{1}{P(0,0)}\text{Tr}_\text{field}\bigg[\left(\Pi(\hat{r}_1^2)+\Pi(\hat{r}_2^2)+\Pi(\hat{r}_3^2)+\Pi(\hat{r}_4^2)+\Pi(\hat{r}_1\hat{r}_3)+\Pi(\hat{r}_1\hat{r}_4)+\Pi(\hat{r}_2\hat{r}_3)+\Pi(\hat{r}_3\hat{r}_4)+\Pi(\hat{r}_2\hat{r}_4)\right)\\
&\qquad \qquad \qquad \qquad \qquad \rho_5\left(\Pi(\hat{r}_1^2)+\Pi(\hat{r}_2^2)+\Pi(\hat{r}_3^2)+\Pi(\hat{r}_4^2)+\Pi(\hat{r}_1\hat{r}_3)+\Pi(\hat{r}_1\hat{r}_4)+\Pi(\hat{r}_2\hat{r}_3)+\Pi(\hat{r}_3\hat{r}_4)+\Pi(\hat{r}_2\hat{r}_4)\right)\bigg],\\
\end{split}
\end{equation}
\end{widetext}

The process can now be repeated from Eq.~\eqref{eq:rhoemitters}, where we replace $\rho_\text{emitters}$ with one of the four possible resulting qubit density matrices, $\rho_{(i,j)}$, depending on the photon-detection outcome.

\bibliography{Papers}{}

\end{document}